\documentclass[twocolumn,showpacs,showkeys,preprintnumbers,amsmath,amssymb,fleqn]{revtex4-2}

\usepackage{graphicx}
\usepackage{dcolumn}
\usepackage{bm}
\begin{document}

\title{Variation in whispering gallery mode of resonance in a trapped and levitated liquid micro cavity}

\author{S. M. Iftiquar}
\affiliation{SPMS-PAP, Nanyang Technological University,21-Nanyang Link, Singapore 637371 }
\affiliation{College of information and Communications Engineering, Sunkyunkwan University, Suwon, South Korea }
\email{smiftiquar@gmail.com}

\date{\today}

\begin{abstract}
Whispering gallery mode (WGM) of resonance occurs when a traveling wave faces grazing reflection and confinement at the inner surface of a spherical cavity. Such a resonance was observed with micrometer sized liquid droplet and with light of sub-micron wavelength. The resonance occurs when specific boundary conditions are fulfilled, following a relation between optical path length (that is related to radius of the droplet) and resonating wavelength. Therefore, a change in WGM wavelength will be related to dimension of the droplet. We observed a shift in such a resonance spectra. For a smaller droplet the observed blue shift in the WGM were 1.5, 0.7, 3.7 nm. Following the resonance condition, it was estimated that such a shift corresponds to a reduction in radius of the droplet by 1.3, 0.6, 3.3 nm respectively. This was obtained with a droplet of radius about 600 nm. The droplet was created from a solution of glycerol, methanol and rhodamine 6G dye, and was trapped and levitated in a modified Paul trap. The WGMs were created by optically exciting the dye material from an external 532 nm cw laser beam. A shift in the WGM was observed with time, during a gradual increase in power of the excitation laser, and a reason for such a shift was thought to be thermal evaporation of the liquid from the droplet. For a larger droplet an initial 0.1 nm thermal expansion was also estimated because initially a red-shift of the WGM was observed, probably because of thermal expansion, which was negligible for a smaller droplet. For the smaller droplet, the estimated rate of change of WGM with radius, was 1.129. For larger droplet, this rate is lower.

\end{abstract}


\maketitle

\section{Introduction}

Experiments with liquid droplet has a long history. Robert A. Millikan used oil drops to measure charge of an electron. In that experiment variation in velocity of the droplets were measured by altering its charge content and changing electric potential to the electrodes through which they fall under gravity. The controlled change in velocity was the key element of that experiment. Later it was found that radiation pressure force can also change velocity of tiny liquid drop [1].  According to Mie scattering this force depends upon size parameter ($X$), where 

\begin{equation}{\label{eqn 1}}
	X = \frac{2 \pi r}{\lambda}
\end{equation}

with $r$ and $\lambda$ are droplet radius and wavelength of radiation respectively [2]. By using this pressure the size of a dielectric sphere was estimated with  1 part in $10^{5}$ accuracy [2].

\subsection{Microdrop}
A spherical droplet can work as an optical cavity where light can confine at its inner surface by total internal reflection (TIR). Based on this, simple lasers were obtained [2-9]. The advantage of a micro cavity is that the coherence length of the light can easily be larger than its round trip path length. Additionally, because of the TIR, the loss of light is expected to be low so a high quality factor or Q factor is achievable [9].  However, it was reported that a droplet laser have a low cavity Q factor because of a presence of quantum dots or dye within it [4].

\subsection{Solidstate ring resonator}

The interest in micrometer sized laser cavity has grown with time. A tiny spherical laser was successfully built in the year 1961 [10]. A microsphere laser was reported to have an advantageously low threshold [11]. A micro ring resonator and laser [12] also works based on whispering gallery mode (WGM) resonance and has been used as sensitive detectors. In such a system a change in temperature or piezo-electric effect can also alter the cavity length, thereby altering the resonance condition. A solid state WGM resonator is an attractive alternative to liquid droplet because of its stability and easy integration on a chip. However, as thermal expansion is an universal property of material, although different material have different coefficient of expansion, therefore we investigated effect of change of cavity by observing WGM mode and estimated the dimensional change than can be a reason for the change in WGM. Here we report observation of  variation in WGM with micro drop when its size changes by few nanometer or even sub-nanometer.

\subsection{Condition of WGM formation}

The assumed condition in which WGM are formed in a micro drop is that the optical path length between two successive reflections should be integral multiple of wavelength. Fluorescence light emitted by dye molecules within the droplet, is uniformly distributed in all possible directions. Out of them a resonating mode can be formed when the above condition is achieved. If $r$ is radius of curvature, $ y $ is length of the chord and $ \lambda $ is wavelength of light, as in Fig 1(a), then the resonance condition is

\begin{equation} {\label{eqn 2}}
	y = m \lambda
\end{equation}

 if  $m$ is an integer.

\begin{figure}
	\centering
	 \includegraphics[width=2cm]{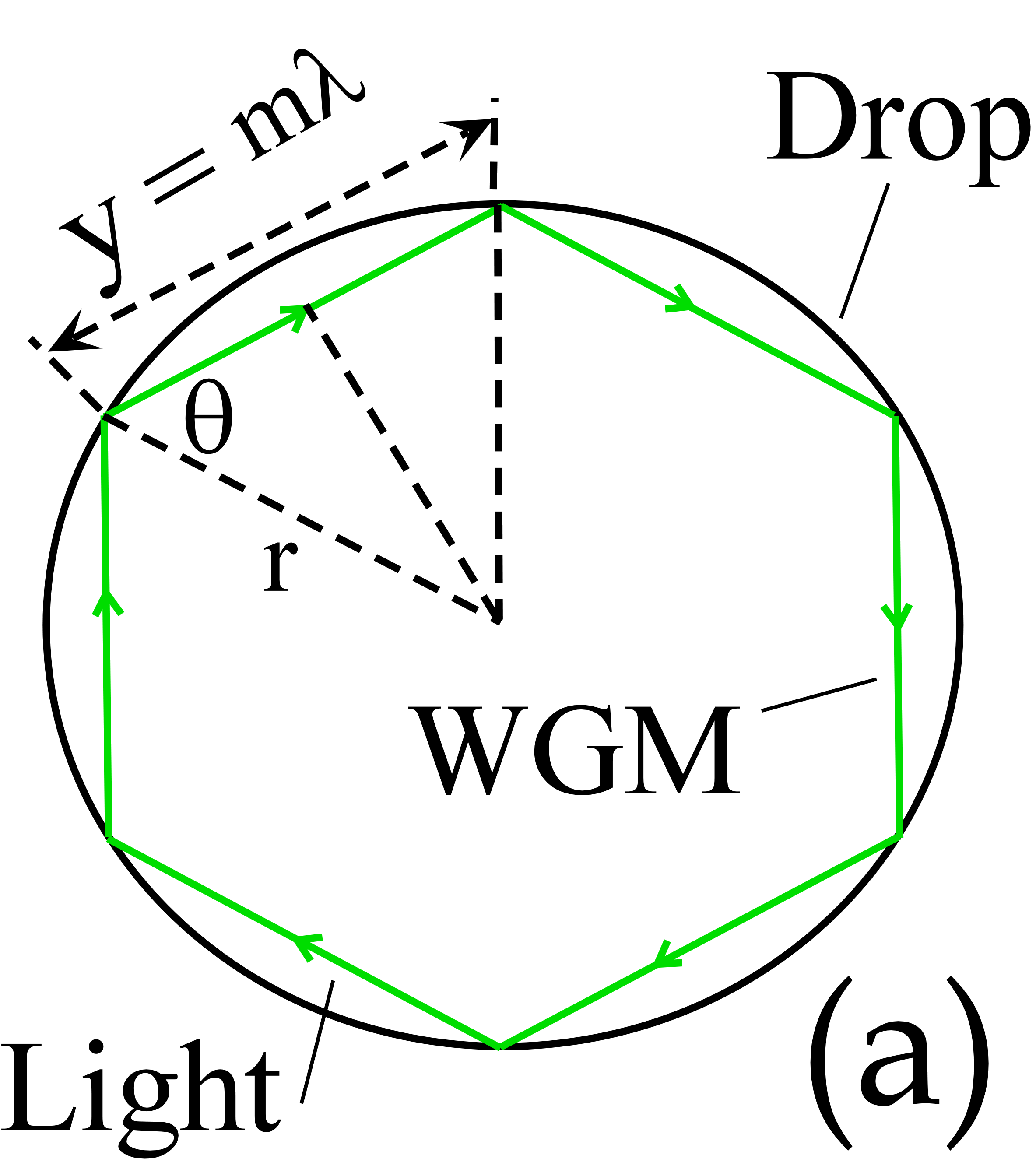} 
	\includegraphics[width=3.5cm]{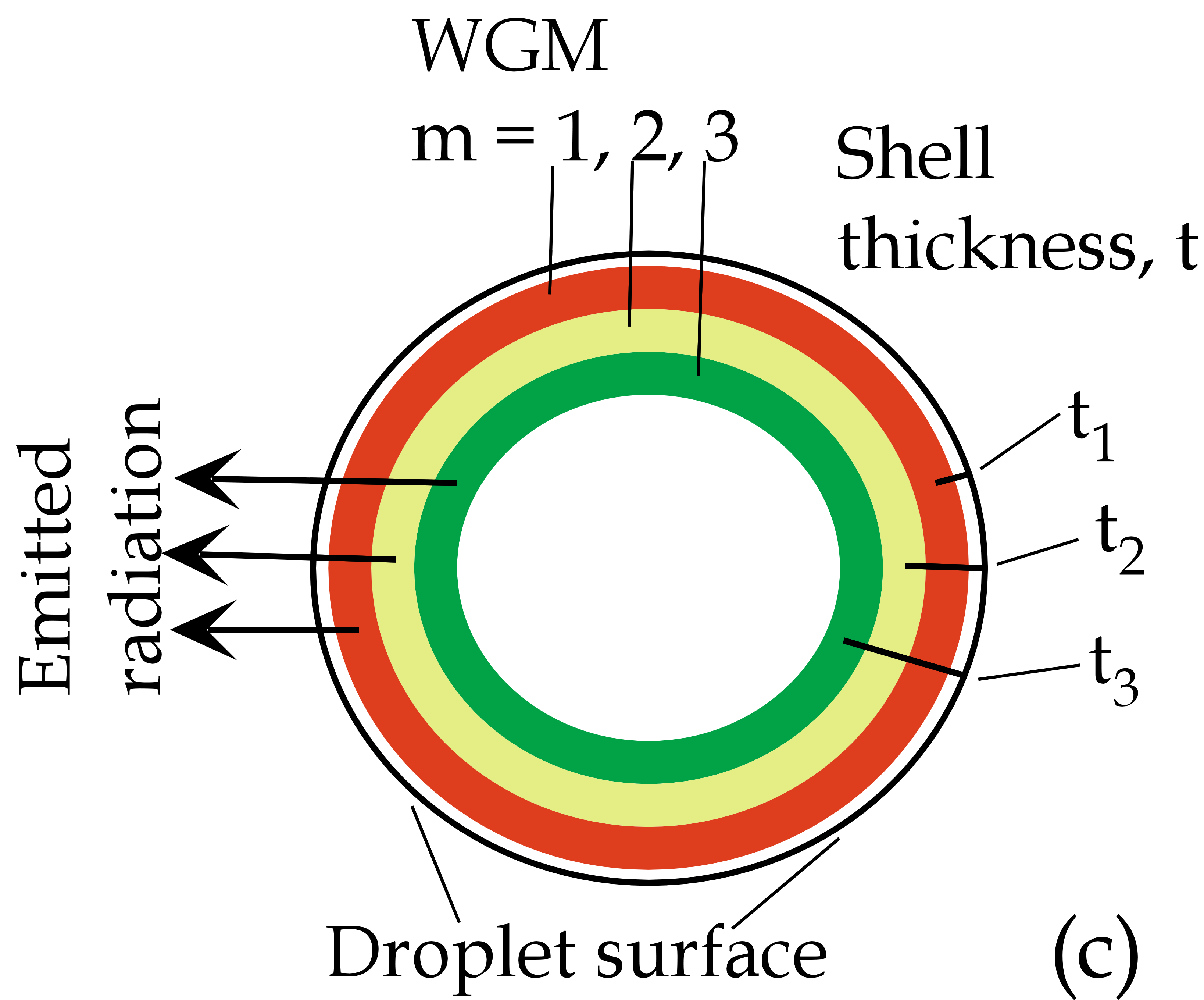}    
	\includegraphics[width=6.5cm]{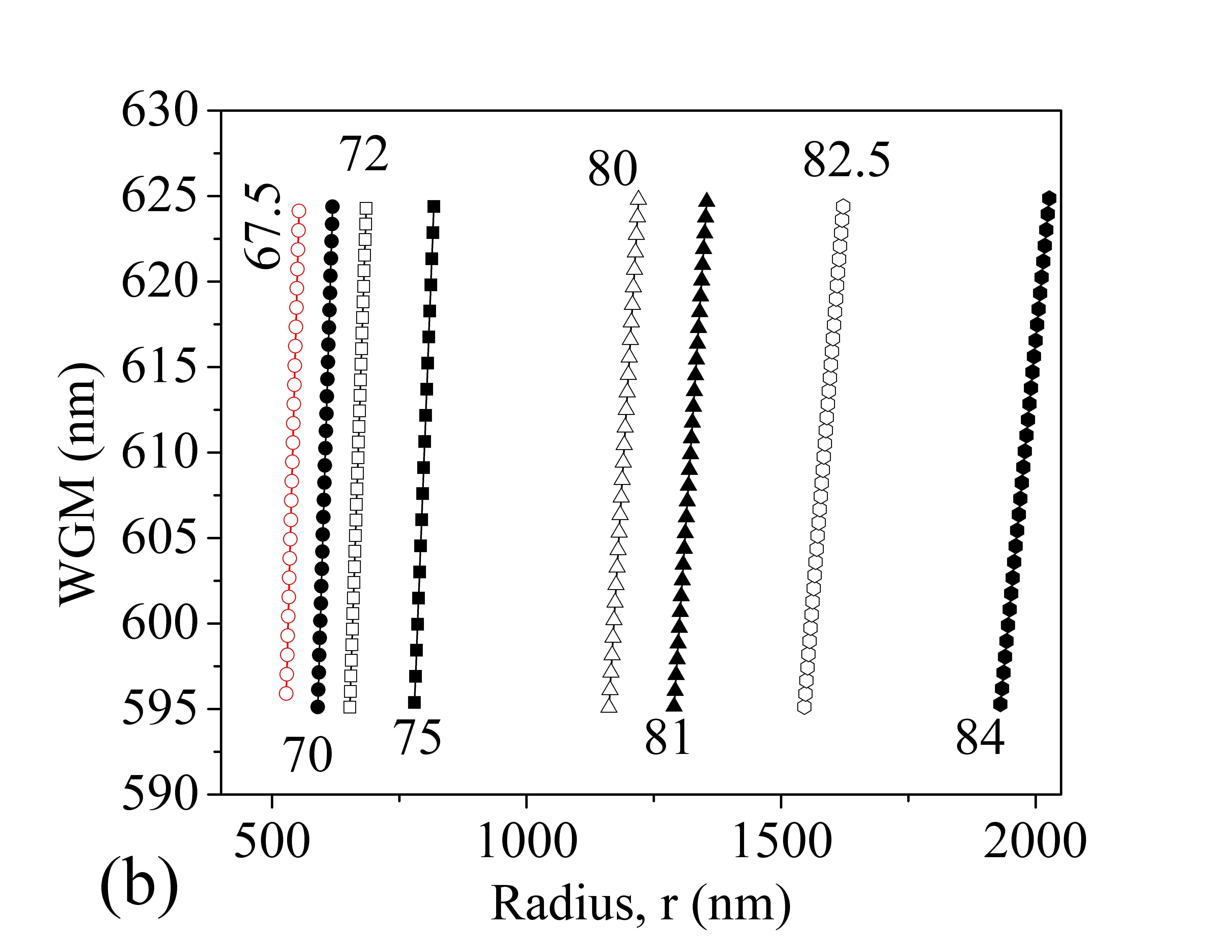}
	\caption[(a)]{Schematic diagram with a circular cross section of a droplet, demonstrating the following, (a) Geometry of a WGM resonance condition. (b) Calculated variation of the WGM for various $r, \theta$, with $m=1$. The parametric values of $\theta$ are indicated as the numbers close to respective traces.  (c) Absorbing path length $t_{1}, t_{2}, t_{3}$ that the corresponding WGM with $m = 1,2,3$ have to travel before radiation into air. }
	\label{fig:fig1}
\end{figure}

According to Fig 1(a), the geometric relation is, 

\begin{equation} {\label{eqn 3}}
	 m \lambda = 2 \mu r \cos \theta 
\end{equation}

Here $ \mu $ is refractive index, so $ 2\mu r \cos \theta $ is the optical path length corresponding to $y$, and $ \theta  $ is the angle of incidence. The expression ( \ref{eqn 3} ) can be compared to Mie scattering parameter $X$. The $ 2 \pi r$ is circumference. If WGM exists very close to the inner surface of the droplet (with $m=1$), and there are $n$ number of reflections for a round trip, then the $ 2 \pi r $ can approximately be the total distance traveled by light in the round trip, and will be approximately equal to $ny$. Then the $ 2 \pi r / \lambda $ will be close to an integer, or,

\begin{equation} {\label{eqn 4}}
 \frac{2 \pi r}{\lambda} \approx \frac{2n\mu r\cos\theta}{\lambda} = nm = n
\end{equation}

Now we define $R_{\lambda} $ , from equation ( \ref{eqn 3}) as

\begin{equation} {\label{eqn 5}}
	R_{\lambda } = \frac{2\mu \cos \theta}{m}
\end{equation}

which is the rate of change of $\lambda$ of WGM with radius $r$.

\subsection{Rate of change of $ \lambda$ with $r$}

Looking into the expression ( \ref{eqn 3} ), it appears counter intuitive that the  $R_\lambda$ will be higher for a smaller droplet than that with a larger one, as $ \lambda$ is proportional to  $ r$. But $R _{\lambda}$ may take different values for different $r$, Fig. 1(b). This is because in order to maintain the resonance condition, the $m$ and $\theta$ have to change with $r$ as well. Here we have a fixed visible wavelength from 595 nm to 625 nm.

The calculation shows that with an increase in droplet size, the $m$ (not shown here) as well as $\theta$ changes. If we keep $m=1$ and unchanged, for example, then the increase in $\theta $ happens for a significantly larger droplet (Fig. 1(b)). The increased $\theta $ makes the $R_{\lambda}$ reduce, as per expression (\ref{eqn 5}). On the other hand as $m = 1$ is the minimum attainable value, so increasing $m$ can reduce the $R_{\lambda}$ further. Both of these happen for larger droplet. Therefore, with a smaller droplet the $R_{\lambda}$ is higher. 

Analyzing the calculated results can make it clearer. Fig. 1(b) shows a few variations in the WGM, for $ m=1 $. From this, the estimated $R_{\lambda}$ are given in Table-I. It shows that the $R_{\lambda}$ is higher for a smaller droplet.

 \tablename{ I. Few selected WGM from droplet cavity, as obtained by using expression (\ref{eqn 3}), with m = 1,2 and the angle $\theta < 83^{o}$. Here $r^{\prime}$ indicates the range of values of the droplet radius.}

\begin{tabular}
	{|c|c|c|c|c|c|c|c|}  
	\hline
	\rule[-1ex]{0pt}{2.5ex}  $m$ & $ \theta^{o}$ & $r^{\prime}$ (nm)& $R_{\lambda}$ &  $m$ & $ \theta^{o}$ & $r^{\prime}$ (nm)& $R_{\lambda}$ \\
	\hline
	\rule[-1ex]{0pt}{2.5ex} 1 & 67.5  & 528-553 & 1.129 & 2  & 45.0  & 571-599 & 1.043\\
	\hline
	\rule[-1ex]{0pt}{2.5ex} 1 & 70.0 & 590-619 & 1.009 &  2 &  60.0 & 807-847 & 0.737 \\
	\hline
	\rule[-1ex]{0pt}{2.5ex} 1 & 72.0 & 653-685 & 0.911 &  2 &  70.0 & 1180-1239 & 0.504 \\
	\hline
	\rule[-1ex]{0pt}{2.5ex} 1 & 75.0 & 780-818 & 0.763 &  2 &  72.0 &  1306-1371 & 0.456 \\
	\hline
	\rule[-1ex]{0pt}{2.5ex} 1 & 80.0 & 1162-1220 & 0.512 & 2  & 75.0  &  1560-1637 & 0.382 \\
	\hline
	\rule[-1ex]{0pt}{2.5ex} 1 & 81.0 & 1290-1354 & 0.461 &  2 &  80.0 & 2324-2440 & 0.256 \\
	\hline
	\rule[-1ex]{0pt}{2.5ex} 1 & 82.5 & 1546-1623 & 0.385  & 2  &  81.0 & 2580-2709 & 0.2331 \\
	\hline
	\rule[-1ex]{0pt}{2.5ex} 1 & 84.0 & 1931-2027 & 0.308 &  2 &  82.5 &  3092-3247 & 0.192 \\
	\hline
\end{tabular}

In addition to the above Fig. 1(b) and Table-I, other results (not given here) shows that there are many more WGM peaks theoretically possible for a larger $r$. Therefore, for a large droplet, identifying the peaks to a specific ($m,\theta$) is not always simple. So we assume the following.

Within the cavity, the chord length ($y$) between two successive TIR have to be an integral multiple of $\lambda$. Higher this chord length is higher is the values of $m$ for a particular $\lambda$, but this mode with higher $m$, on the average, have to travel along a shell that is deeper within the droplet, Fig 1(c). Therefore, the  intensity of the radiation will be weaker for a higher $m$, because of more light absorption at the thicker shell. Based on his, the most intense of the WGM appear from the mode that is closest to the surface. This corresponds to the least value of $m$, i.e. $m=1$ and subsequently weaker peaks correspond to $m=2, 3$ etc.   

\section{Experiment}

A modified Paul trap was used to trap and levitate a negatively charged microdrop, as shown in Fig. 2. The electrodes are conically shaped and were centrally hollow, Fig. 2(a),(b), while the electrospray needle (not shown here) placed above it. They are arranged vertically so that droplets can pass through its holes and be captured by the electrodes, Fig. 2(c). Initially, multiple droplets were captured, then one of those was selectively trapped and levitated for a prolonged investigation, Fig 2(c). This investigation was carried out with rhodamine 6G (Rh6G) dye, with $60\%$ glycerol and $40\%$ methanol and 20mg/ml Rh6G. A  cw laser of $532$ nm wavelength, variable output optical power (0 to 300 $\mu$m), was used to optically excite the dye. Light for the CCD camera and spectrometer were collected perpendicular to the excitation beam direction [4,6]. The spectra were recorded by Ocean Optics spectrometer.

\begin{figure}
	\centering
	\includegraphics[width=3.5cm]{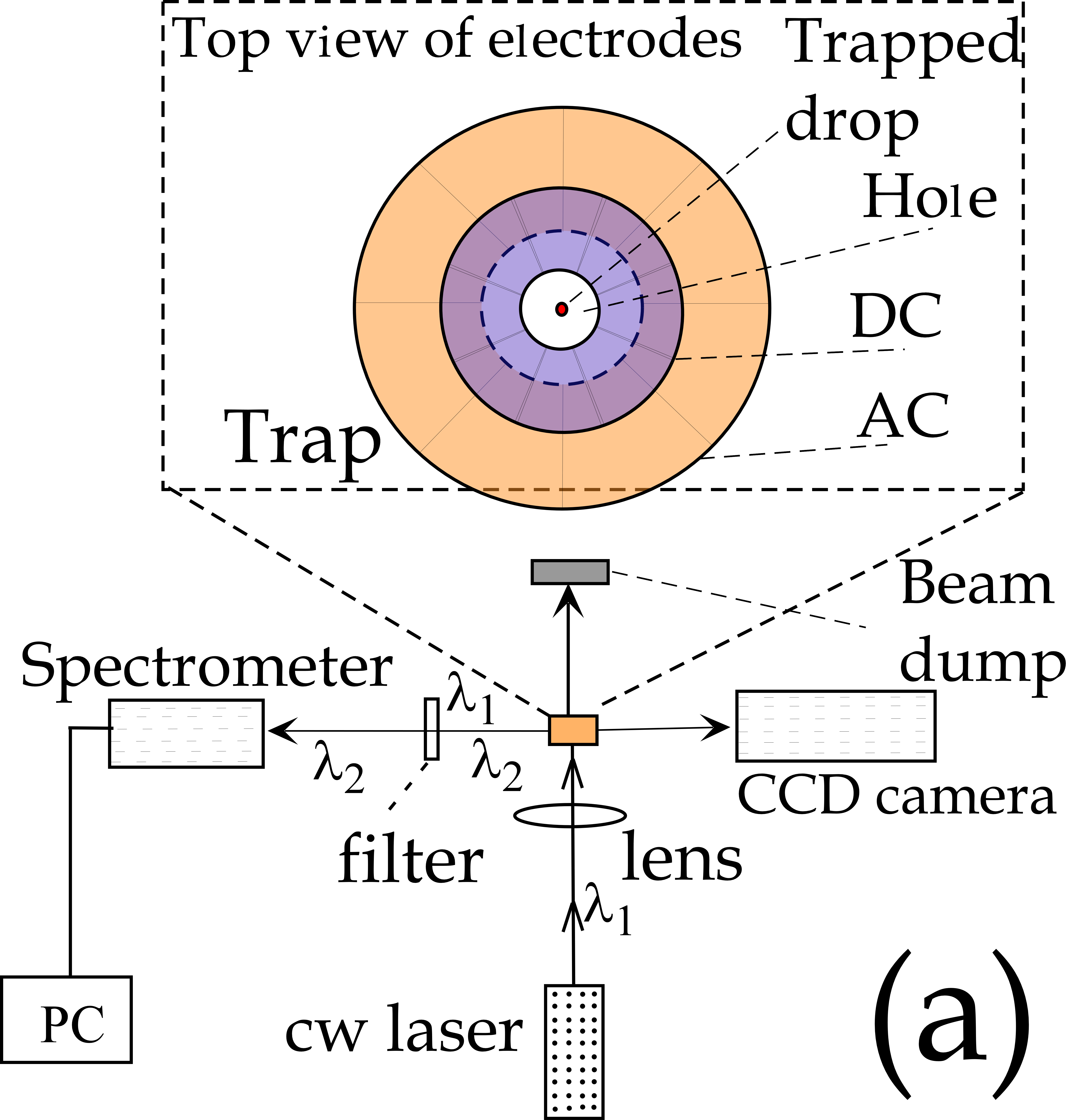} 
	\includegraphics[width=3.5cm]{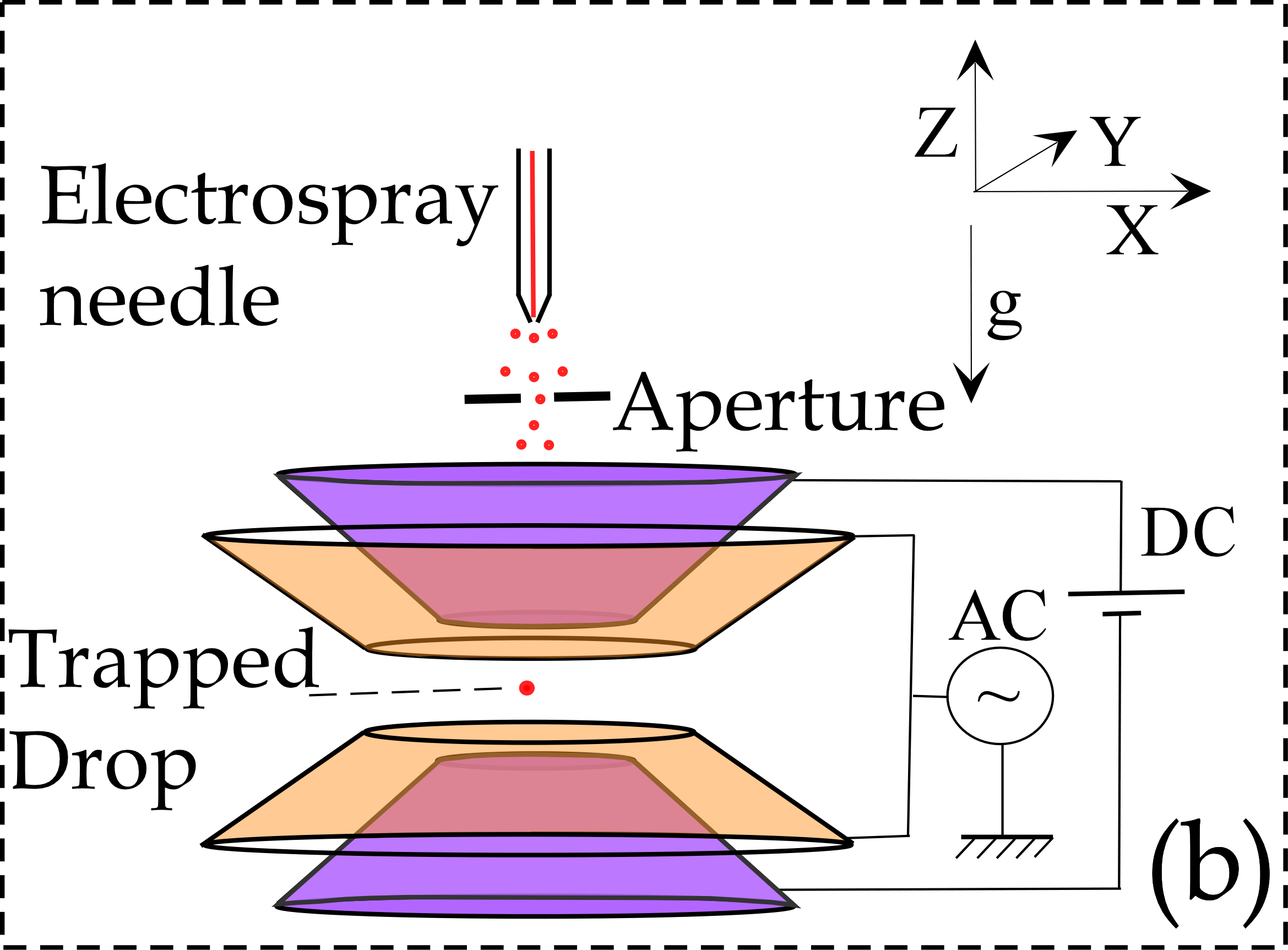}    
	\includegraphics[width=3.5cm]{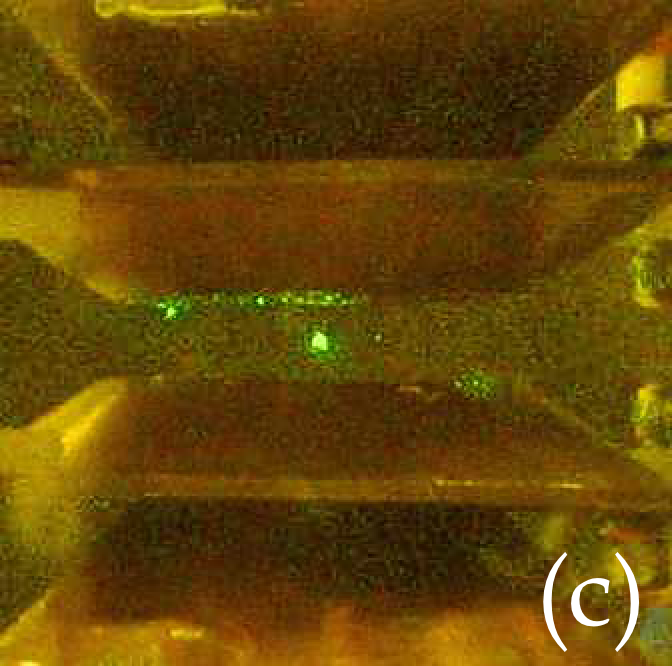}
	\includegraphics[width=3.5cm]{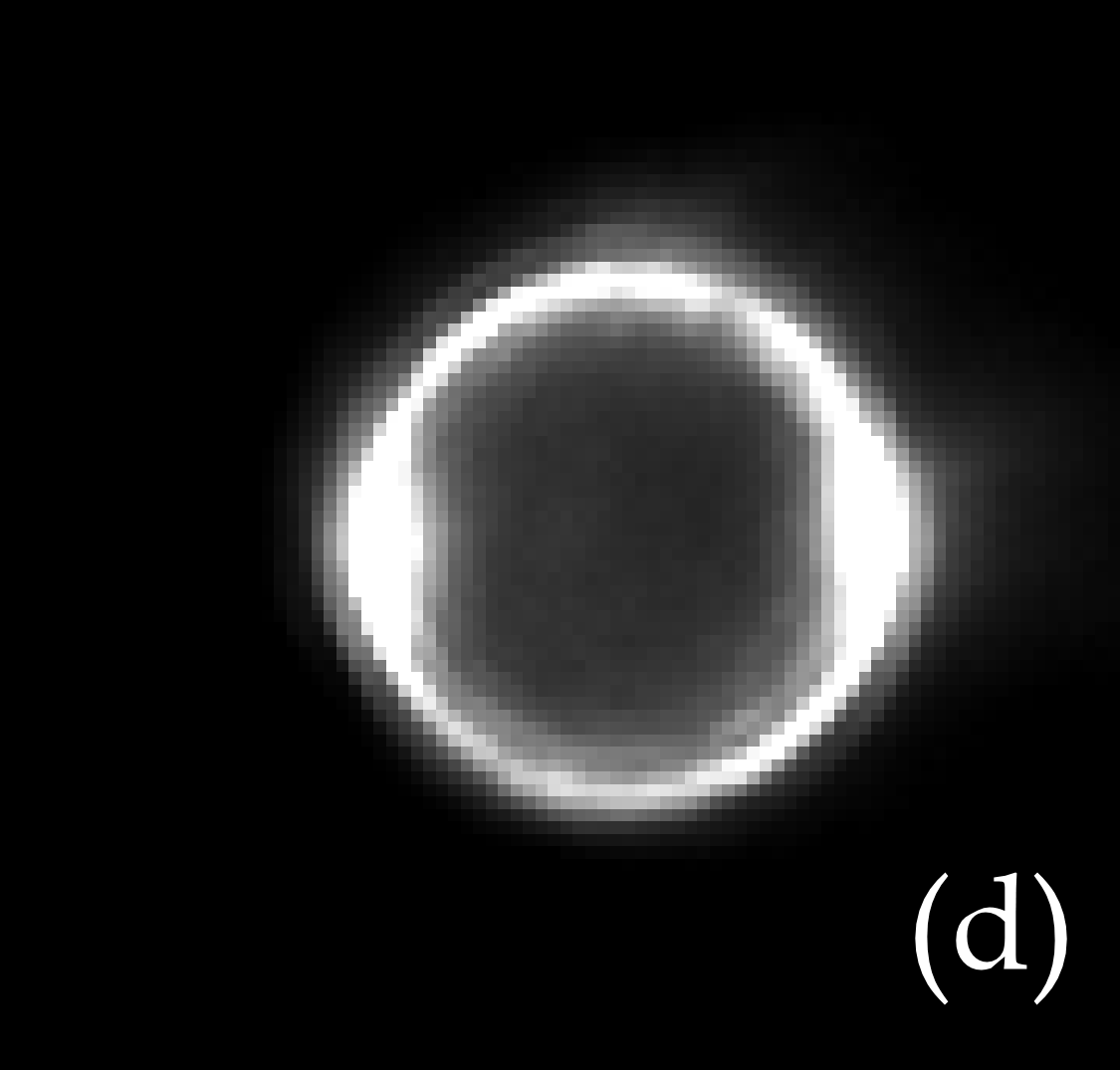}
	\caption[(a)]{Schematic diagram of a modified Paul trap. (a) Top view of the set up, (b) side view of trapping electrodes. (c) Image of the trapping electrodes when a droplet was trapped and illuminated by the 532 nm laser light. (d) CCD image of a droplet that was optically excited by the 532 nm laser and observed with a 532 nm cutoff filter. So the image is due to the fluorescence light.  }
	\label{fig:fig2}
\end{figure}

\section{Results and Discussions}

Figure 2(d) shows an image of a droplet, taken with a 532 nm filter to cut-off the excitation light. Therefore, the bright ring corresponds to WGM resonance. Figure 3 shows a set of four WGM spectra from a droplet of 600 nm radius. The Fig. 3(a) to (d) were recorded while the excitation laser power was increased in steps and slowly. The numbers in the figures indicate the peak positions in nm.  It can be noticed that only a single peak was initially visible, corresponding to the $m=1$ mode. However, with an increase in the laser intensity two side lobes appeared. As the measurement of the droplet size with the CCD image was not very accurate, so we estimated the $r$ again based on the theoretical calculation mentioned above. In the Fig.3(a) the $\lambda = 609.9$ nm, so the radius, say $r_{1}, r_{1}=540.2$ nm. Similarly, from the second spectra, Fig. (b) the radius $r_{2} = 538.9$ nm, that is a decrease in radius ($r_{1} - r_{2}$) by $1.3$ nm. From Fig. 3(c), $r_{3} = 538.3$ nm, and with a decrease $r_{2} - r_{3}=0.6$ nm. And from Fig. 3(d) $r_{4}=535.0$ nm, with $r_{3} - r_{4}=3.3$ nm. It shows that the radius of the droplet continued to decrease. It was because of thermal evaporation of the droplet liquid. 

In Fig. 3(c), (d) there are two additional peaks visible. In Fig. 3(c), these are at $\lambda = 595.6$ nm and $621.4$ nm. We identify these peaks are for $m=2$. The $\lambda =595.6$ nm corresponds to radius $r_{5}= 571$ nm, and  the $\lambda =621.4$ nm corresponds to radius $r_{6}= 596$ nm. It may appear a bit strange but probably these are the semi-axes of tri-axial ellipsoidal droplet. As the droplet continued to absorb laser light, its temperature is expected to increase. The change from spherical shape to the ellipsoid form seems to be initiated by this rise in temperature and gravitation pull. Similarly the other higher order WGM ($m=2$) in Fig 3(d), $\lambda = 591.9$ nm gives $r_{7}=570$ nm and for the $\lambda =618.9 $ nm the $ r_{8}=593.5$ nm. So it is clear from the combination of $r_{5}= 571, r_{7}= 570$ nm $r_{6}=586, r_{8}= 593.5$ nm and  that the dimension of the three semi-axes decreased from that of the Fig. 3(c).

\begin{figure}
	\centering
	\includegraphics[width=9cm]{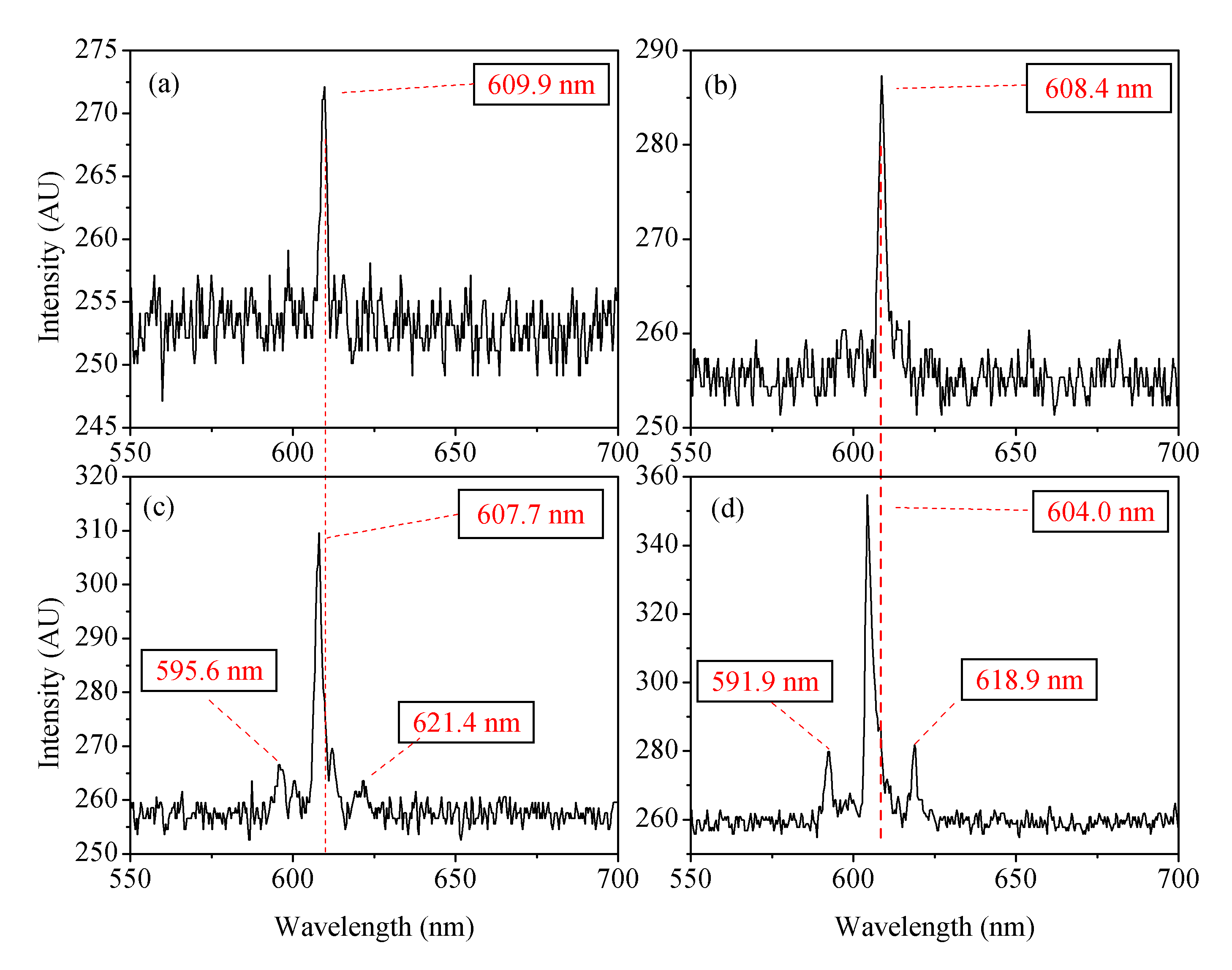} 
	\caption[]{WGM spectra of a droplet with approximate $r = 600$ nm. Fig. (a) to (d) were recorded for the same drop but with an increased intensity of excitation cw laser.  }
	\label{fig:fig3}
\end{figure}

A similar description can be given for a larger droplet. But in this case, the number of observed peaks increase, as shown in Fig. 4, for $r=1800$ nm.

\begin{figure}
	\includegraphics[width=9cm]{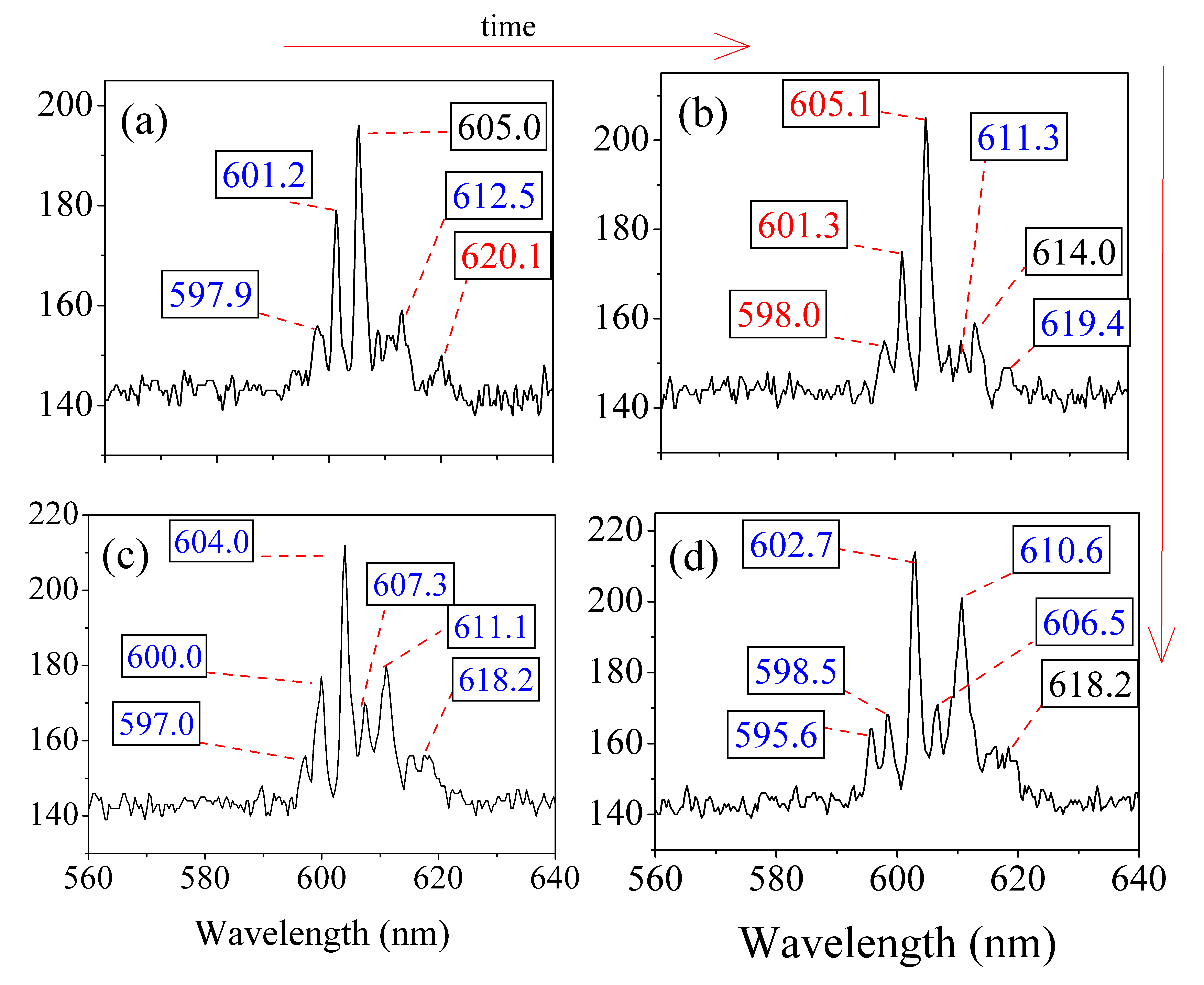} 
	\caption[]{WGM spectra of a droplet with approximate $r = 1.8 \mu$m. As like Fig. 3, here also the Fig. (a) to (i) were recorded for the same drop but with an increased intensity of excitation cw laser. Numbers within the traces indicate spectral peak position in nm.  }
	\label{fig:fig4}
\end{figure}

The most intense peak at $\lambda = 605$ nm is for $m=1$  and $r = 1572$ nm. Then $\lambda = 601.2$ nm is for $m=2$. The experimental error for measuring $r$ by the CCD camera is 1 pixel or 729 nm. On the other hand, the spectroscopic method gives a sub-nanometer precision. In Fig. 4(a),(b) the WGM for $m=1$, shifts from  $\lambda = 605.0$ to $605.1$ nm indicates an increase in droplet size by $0.1$ nm due to rise in its temperature. This is because with time more light energy will be absorbed by the droplet than the loss by radiation or liquid evaporation, leading to rise in its temperature, therefore the thermal expansion and observed red shift of the $m=1$ mode. With time, due to steady evaporation of the liquid, its size decreases leading to a blue shift of the WGM peaks. The additional peaks can similarly be identified and the related radius or semi-axes of the droplet can be estimated. The WGM for m=2 is  shown in Table-II. It can be noticed that for the larger droplet the initial red shift takes place due to its thermal expansion, but later its radius reduces due to evaporation, leading to a noticeable blue shift of the peaks. This expansion was not noticed for a smaller droplet, probably because the size reduction was below the detection limit.

\tablename{ II. Selected WGM from droplet of radius 1800 nm. $\lambda_{1}, \lambda_{2}$ are the wavelengths for $m=1,2$. The data is taken from Fig. 4. Reference sub-figure numbers are given. Here $r$ is estimated radius from $m=1$ mode. }

\begin{tabular}
	{|c|c|c|c|c|c|}  
	\hline
	\rule[-1ex]{0pt}{2.5ex}  sub-figure number & $m$ & $ \lambda_{1}$ (nm) &  $r$ (nm)  &  Fig &  $ \lambda_{2}$ (nm)  \\
	\hline
	\rule[-1ex]{0pt}{2.5ex}  a &   1   &   605.0         &  1572.0    & a  & 601.2   \\
	\hline
	\rule[-1ex]{0pt}{2.5ex}  b &   1   &  605.1          &  1572.1    &  b & 601.3   \\
	\hline
	\rule[-1ex]{0pt}{2.5ex}  c &   1   &   604.0         &   1569.0   &  c & 600  \\
	\hline
	\rule[-1ex]{0pt}{2.5ex}  d &   1   &   602.7         &   1566.0  & d &  598.5   \\
	\hline
\end{tabular}
\ 
The values of $r$ that are estimated from Fig. 3 and Fig. 4, may appear inconsistent or arbitrary, as the WGM  for $m=1$ in these two situations are very close to each other, yet the estimated $r$ varies significantly. However, looking into the Fig. 1(b) one may come to the conclusion that one particular peak may be due to a large range of possible $r$. For example, with $m=1$ and  $\lambda=600$ nm, the $r$ can be as low as 500 nm or as high as 2000 nm. This situation is true not only for $\lambda=600$ nm but also for the full range of $\lambda$. Therefore, an initial approximate measurement of size of the droplet is necessary to select an appropriate range of values. After that an accurate estimate of its size from the WGM is possible. Meaning, although the method of using WGM to estimate $r$ gives a high precision, but its range of applicability is limited; increased precision comes at the expense of a reduced range.
Furthermore, it may also be noticed that the wavelength for WGM spectral peak depends on size of droplet as well, Fig. 1(b). Although it is known that dye concentration determines the fluorescence wavelength, but the fluorescence has a broader spectral range, while the WGM is narrower and wavelength selective, signifying the WGM as directly related to a specific cavity configuration. So the peak position of the mode can vary greatly even with a small change in $r$, Fig 1(b). For example, the $\lambda$ increases linearly from 595 to 625 nm when $r$ changes from $528$ to $553$ nm as well as from 9634 to 10000 nm.

\section{Conclusion}
In conclusion, we have observed a shift in WGM resonance of a micro drop optical cavity. A red shift corresponds to an increase in droplet size while the blue shift of the WGM corresponds to decrease in its size. Thermal expansion can raise the droplet size while evaporation will lead to its reduction. For a smaller droplet the thermal expansion was not clearly visible as its pressure due to surface tension will be higher, the amount of liquid content is lower and may also be out of detection range. The red shift of the WGM for $m=1$ was not observed for the small drop ($r=600 $nm). On the other hand the larger drop shows a red shift in WGM.

%



\section{References}
\begin{thebibliography}{12}

\bibitem [1]{ref 1}
A. Ashkin, J.M. Dziedzic, Optical Levitation of Liquid Drops by Radiation Pressure, Science, 187 (1975) 1073-1075. 

\bibitem [2] {ref 2} 
A. Ashkin, J.M. Dziedzic, Observation of Resonances in the Radiation Pressure on Dielectric Spheres, Physical Review Letters, 38 (1977) 1351-1354. 

\bibitem [3] {ref 3} 
D. McGloin, Droplet lasers: a review of current progress, Reports on Progress in Physics, 80 (2017) 054402. 

\bibitem [4] {ref 4} 
S.M. Iftiquar, Y. Wijaya, R. Dumke, Characterization of (CdSe)ZnS core-shell quantum dots in microdrop laser cavity, Optics and Photonics Letters, 04 (2011) 1-10.  

\bibitem [5] {ref 5} 
S.M. Iftiquar, Levitated microdrop quantum dot and dye laser in a modified Paul trap, J Opt (India), 41 (2012) 110-113 .  

\bibitem [6] {ref 6} 
S.M. Iftiquar, H. Zilay, Investigation of variation in fluorescence intensity from rhodamine 6G dye in a trapped and levitated liquid micro-drop, Optik, 291 (2023) 171351.  

\bibitem [7] {ref 7} 
H. Zhang, P. Palit, Y. Liu, S. Vaziri, Y. Sun, Reconfigurable Integrated Optofluidic Droplet Laser Arrays, ACS applied materials and interfaces, 12 (2020) 26936-26942.  

\bibitem [8] {ref 8} 
X. Chen, L. Fu, Q. Lu, X. Wu, S. Xie, Packaged Droplet Microresonator for Thermal Sensing with High Sensitivity, Sensors (Basel, Switzerland), 18 (2018) 3881. 

\bibitem [9] {ref 9} 
A. Giorgini, S. Avino, P. Malara, P. De Natale, G. Gagliardi, Liquid Droplet Microresonators, Sensors (Basel, Switzerland), 19 (2019) 473.  

\bibitem [10] {ref 10}
 C.G.B. Garrett, W. Kaiser, W.L. Bond, Stimulated Emission into Optical Whispering Modes of Spheres, Physical Review, 124 (1961) 1807-1809. 

\bibitem[11] {ref 11}
 V.V. Sandoghdar, F. Treussart, J. Hare, V.V. Lefèvre-Seguin, J. Raimond, S. Haroche, Very low threshold whispering-gallery-mode microsphere laser, Physical review. A, Atomic, molecular, and optical physics, 54 (1996) R1777-r1780.

\bibitem[12]{ref 12} 
 S.L. McCall, A.F.J. Levi, R.E. Slusher, S.J. Pearton, R.A. Logan, Whispering gallery mode microdisk lasers, Applied Physics Letters, 60 (1992) 289-291.  

\end{thebibliography}
\end{document}